\documentclass{article}

\usepackage[final]{ml4eng_2020}

\usepackage[utf8]{inputenc} 
\usepackage[T1]{fontenc}    
\usepackage{url}            
\usepackage{booktabs}       
\usepackage{amsfonts}       
\usepackage{nicefrac}       
\usepackage{microtype}      
\usepackage{bm}
\usepackage{enumitem}
\usepackage{graphicx}
\usepackage{amsmath}
\usepackage{subfigure}
\usepackage[numbers, sort]{natbib}

\usepackage[colorlinks=true,bookmarksopen,pdfsubject={algorithms},linkcolor={blue}, anchorcolor={black}, citecolor={blue}, filecolor={magenta}, menucolor={black}, pagecolor={red},backref=none,urlcolor={blue}]{hyperref}
            
\title{Scalable Deep-Learning-Accelerated Topology Optimization for Additively Manufactured Materials}

\author{Sirui Bi \\
   Computational Sciences and Engineering Division  \\
   Oak Ridge National Laboratory, Oak Ridge, TN 37830 \\
  \texttt{bis1@ornl.gov} \\
   \AND
   Jiaxin Zhang \\
  Computer Science and Mathematics Division\\
  Oak Ridge National Laboratory, Oak Ridge, TN 37830 \\
  \texttt{zhangj@ornl.gov} \\
  \AND
   Guannan Zhang \\
   Computer Science and Mathematics Division \\
   Oak Ridge National Laboratory, Oak Ridge, TN 37830 \\
  \texttt{zhangg@ornl.gov} \\
}

\begin{document}

\maketitle
\begin{abstract}

Topology optimization (TO) is a popular and powerful computational approach for designing novel structures, materials, and devices. Two computational challenges have limited the applicability of TO to a variety of industrial applications. First, a TO problem often involves a large number of design variables to guarantee sufficient expressive power. Second, many TO problems require a large number of expensive physical model simulations, and those simulations cannot be parallelized. To address these issues, we propose a general scalable deep-learning (DL) based TO framework, referred to as SDL-TO, which utilizes parallel schemes in high performance computing (HPC) to accelerate the TO process for designing additively manufactured (AM) materials. Unlike the existing studies of DL for TO, our framework accelerates TO by learning the iterative history data and simultaneously training on the mapping between the given design and its gradient. The surrogate gradient is learned by utilizing parallel computing on multiple CPUs incorporated with a distributed DL training on multiple GPUs. The learned TO gradient enables a fast online update scheme instead of an expensive update based on the physical simulator or solver. Using a local sampling strategy, we achieve to reduce the intrinsic high dimensionality of the design space and improve the training accuracy and the scalability of the SDL-TO framework. The method is demonstrated by benchmark examples and AM materials design for heat conduction. The proposed SDL-TO framework shows competitive performance compared to the baseline methods but significantly reduces the computational cost by a speed up of around 8.6x over the standard TO implementation.

\end{abstract}

\section{Introduction}

Topology optimization (TO) has been extensively applied to solving complex engineering design problems with a wide range of industrial applications, e.g. aerospace, building, and automotive. Unfortunately, TO has a major challenge in computational expense because it is an iterative and sequential procedure, which is not straightforward to be palatalized. A typical large-scale TO problem may involve hundreds or even thousands of design iterations, and for each iteration, the physical response needs to be solved to compute the gradient information through sensitivity analysis. To handle large-scale TO problem, which involves millions of design variables \cite{aage2017giga,alexandersen2016large, evgrafov2008large}, the associated computational cost is highly prohibitive and will be larger and larger in the future since the problem are becoming more complex with more details \cite{aage2017giga}. In general, there are several types of methods to improve the computational efficiency, including algorithm improvements \cite{amir2011reducing, li2020accelerated}, e.g., advanced iterative solvers \cite{liao2019triple, ferrari2020new}, parallel computing \cite{aage2017giga,aage2013parallel} including GPU computing \cite{martinez2017gpu,zegard2013toward,xia2017gpu}, and deep learning techniques \cite{qian2020accelerating,sasaki2019topology}, e.g., reparameterization \cite{kallioras2020accelerated,hoyer2019neural}. 

Another typical strategy to improve the efficiency of TO simulations is to build surrogate modeling, which has been developed in a wide variety of contexts and disciplines, see \cite{BTNT12,CCDS13,CCS14,Cohen:2010kz} and the reference therein. Surrogate modeling practice seeks to approximate the response of an original function (the loss function in this work), which is typically computationally expensive, by a cheaper-to-run surrogate. The loss function can then be evaluated by evaluating the surrogate directly without running expensive forward model executions. Compared to conventional TO algorithms, this approach is advantageous that it significantly reduces the number of forward model executions at a desired accuracy and allows evaluating the loss function in parallel. Several methods can be employed to construct the surrogate systems, including least-squares projection \cite{Migliorati:2014ifa,Migliorati:2013kba,Chkifa:2013ww,Migliorati:2014ws}, sparse interpolation \cite{Nobile:2008wf, Nobile:2008dr, Eldred:2008Pb}, and compressive sensing \cite{DO11, MG12, YGX12, YK13, RS14, PHD14}, etc. However, unlike the standard surrogate modeling task, when TO does not require an accurate surrogate model in the entire parameter space, because an optimizer only goes down a single path from the initial state to the final state to find the optimum. Thus, it is wasteful to build a global surrogate to the loss function. On the other hand, as TO is often a high-dimensional optimization problem, it is also challenging to build a global surrogate with sufficient accuracy even with the state-of-the-art neural network models.

In this work, we first propose a general scalable framework integrating deep learning and parallel computing to accelerate the TO process for designing additively manufactured materials. The key idea is to dig into the iterative history data and use deep neural networks (DNN) to learn a surrogate gradient given a specific design. The learning step is achieved by utilizing parallel computing on multiple CPUs and a distributed DNN training on multiple GPUs, incorporated with a novel local sampling strategy, which aims to reduce the dimensionality to capture a low-dimensional search space. The proposed SDL-TO framework is demonstrated by several benchmark examples and compared with the start-of-the-art baseline methods. In the large-scale AM materials design example, SDL-TO significantly saves the computational cost and achieves a speed up of around 8.6x over the standard TO implementation. 

\section{SDL-TO: Scalable Deep Learning Accelerated Topology Optimization}

\begin{figure}[h!]
     \centering
     \includegraphics[width=0.98\textwidth]{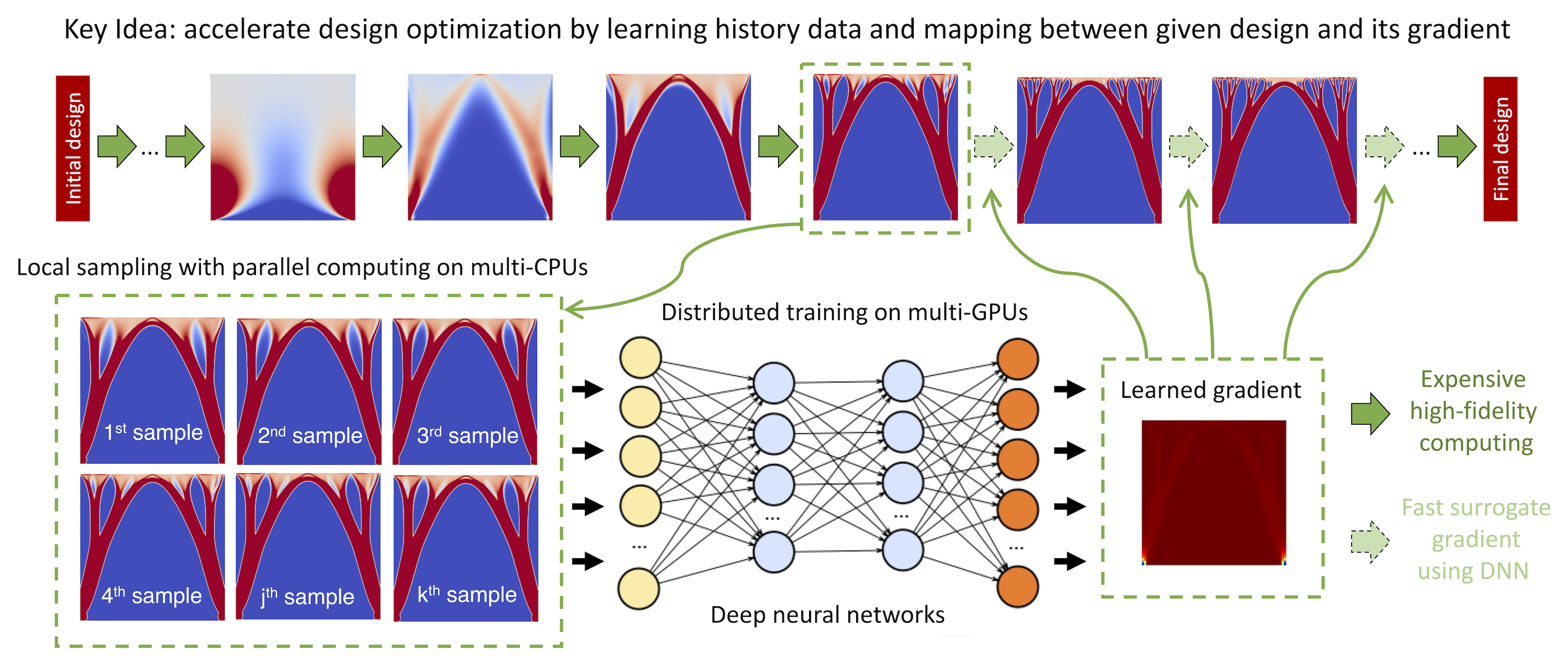}
    \caption{SDL-TO framework to accelerating TO by learning the iterative history data and mapping between given design and its gradient. Using local sampling strategy with distributed computing and training on multiple CPUs and GPUs, the learned surrogate gradient enables an online update instead of expensive computing using a high-fidelity simulator or solver.}
    \label{fig:SDL-TO}
    \vspace{-0.2cm}
\end{figure}

In this paper, we propose a scalable deep learning framework to accelerate TO, but we emphasize that it can be generally applied to a wide range of computational design optimization problems, specifically large-scale, high-dimensional, computationally intensive problems in computational sciences and engineering (CSE). The core idea is using an accurately learned gradient by DNN instead of the true gradient that is iteratively evaluated at each step to accelerate the optimization process. This is achieved by utilizing a local sampling strategy incorporating a scalable implementation on multiple CPUs and GPUs. Figure \ref{fig:SDL-TO} gives an overview of this process. Note that the proposed framework is universal in the sense that it can work with any gradient-based optimization based on computationally intensive physical models or simulators. 

{\bf Topology optimization formulation} \quad TO is a mathematical method that aims to optimize materials layout defined on a design domain with given boundary conditions, loads and volume constraints to minimize structural compliance $C$ (or equivalently the least strain energy). In this work, we use the modified Solid Isotropic Material with Penalization (SIMP) approach \cite{sigmund2007morphology} with density-based approach to TO, where each element $e$ is assigned a density $x_e$ that determines its Young's modulus:
\begin{equation}
    E_e(x_e) = E_{\min} + x_e^p(E_0-E_{\min}),\quad x_e \in [0,1]
\end{equation}
where $E_0$ is the stiffness of the material, $E_{\min}$ is a very small stiffness assigned to void regions to prevent the stiffness matrix becoming singular. The modified SIMP approach differs from the classical SIMP approach \cite{bendsoe1989optimal}, where elements with zero stiffness are avoided by using a small value. The modified mathematical formulation of the TO problem is
\begin{equation}
    \min_{{\bm x}}: C(\bm x) = \mathbf{U}^T \mathbf{K} \mathbf{U} = \sum_{e=1}^N E_e(x_e) \mathbf{u}_e^T \mathbf{k}_0 \mathbf{u}_e, \quad {\rm s.t.} \quad \mathbf{KU=F}, \ V(\bm x)= V_0, \ 0 \le \bm x \le 1
\end{equation}
where $\bm x$ is the vector of design variables, $C$ is the structural compliance, $\mathbf{K}$ is the global stiffness matrix, $\mathbf{U}$ and $\mathbf{F}$ are the global displacement and force vectors respectively, $\mathbf{u}_e$ and $\mathbf{k}_e$ are the element displacement vector and stiffness matrix respectively, $N$ is the number of elements used to discretize the design domain $\Omega$, $V(\bm{x})$ and $V_0$ are the material volume and design domain volume respectively, $\zeta$ is the prescribed volume fraction, and $p$ is the penalization power coefficient (typically $p=3$). 

{\bf TO baseline method} \quad The computationally limiting step in standard TO is the gradient estimation using an adjoint method \cite{bendsoe2013topology}, which typically needs to solve $\mathbf{U=K^{-1}F}$ using the finite element method (FEM). The gradient is inserted into the Method of Moving Asymptotes (MMA) \cite{svanberg1987method} algorithm that is the state-of-the-art optimizer, which has been extensively demonstrated to be versatile and well suited for a wide range TO problems. We implemented this algorithm using \texttt{NumPy} and compare our method with the standard TO process using MMA in a single CPU. Note that our scalable framework aims to accelerate the iterative optimization process and the comparison is based on the same solver and TO algorithm. The improvement of the TO algorithm self, the linear solver, or the physical simulator using parallel computing is out of the scope of this study. 

{\bf Online update with a local sampling strategy} \quad To large-scale TO problems, typically involving high resolution, the design space is so huge that it is very challenging to directly map the given design and its gradient using random samples. It is a non-trivial task to use a global surrogate model to cover the entire high-dimensional space in this case. To address this issue, we propose a local sampling strategy that allows us to capture the optimal low-dimensional gradient direction by generating random samples around the current design with a few history steps before. Since the standard TO algorithm uses gradient descent for a local search, the adjacent design does not show significant changes except for the initial stage. Therefore, we propose a local sampling strategy using local samples that are drawn from a Gaussian distribution 
\begin{equation}
    \tilde { \bm{x}}_{rs}^{j+1} \sim \mathcal{N}(\bm{x}_{\textup{min}}^{j+1} +(\bm{x}_{\textup{max}}^{j+1}-\bm{x}_{\textup{min}}^{j+1})/2 , \bm \sigma) \label{eq:gaussian}
\end{equation}
where $\sigma$ is the standard deviation which is used to control the radius of the local search domain, and the upper and lower bound are defined by 
\begin{equation}
   \tilde { \bm{x}}_{\textup{max}}^{j+1}=\max\left\{ \tilde { \bm{x}}^{(j)},\tilde { \bm{x}}^{(j-1)}, \cdots, \tilde { \bm{x}}^{(j-w)}\right\}, \quad \tilde { \bm{x}}_{\textup{min}}^{j+1}=\min\left\{ \tilde { \bm{x}}^{(j)},\tilde { \bm{x}}^{(j-1)}, \cdots, \tilde { \bm{x}}^{(j-w)}\right\}   \label{eq:bound}
\end{equation}
where $w$ is a lookback window size, which means how many history steps we will look back from the current state $\tilde { \bm{x}}^{j+1}$. As shown in Figure \ref{fig:SDL-TO}, we have a number of designs and their corresponding gradients from these local samples. A deep neural network is trained to map the design and its gradient and we therefore learn a accurate gradient for any specific design within this local search domain. Once we have the learned gradient in hand, we may continue gradient-based optimization without any expensive simulations, e.g., FEM, but a critical question is bought up: {\em how many steps can we move forward to the optimal design?} This is determined by defining a threshold $\lambda^*$ of the cosine distance $d^c$ between current design $\tilde { \bm{x}}^{j+1}$ and future design $\tilde { \bm{x}}^{j+m}$. If $d^c>\lambda^*$, the {\em learning} step consisting of simulation and training will be called, otherwise, the learned surrogate gradient will be used for online next step update of the design. 
\begin{figure}[h!]
     \centering
     \includegraphics[width=0.98\textwidth]{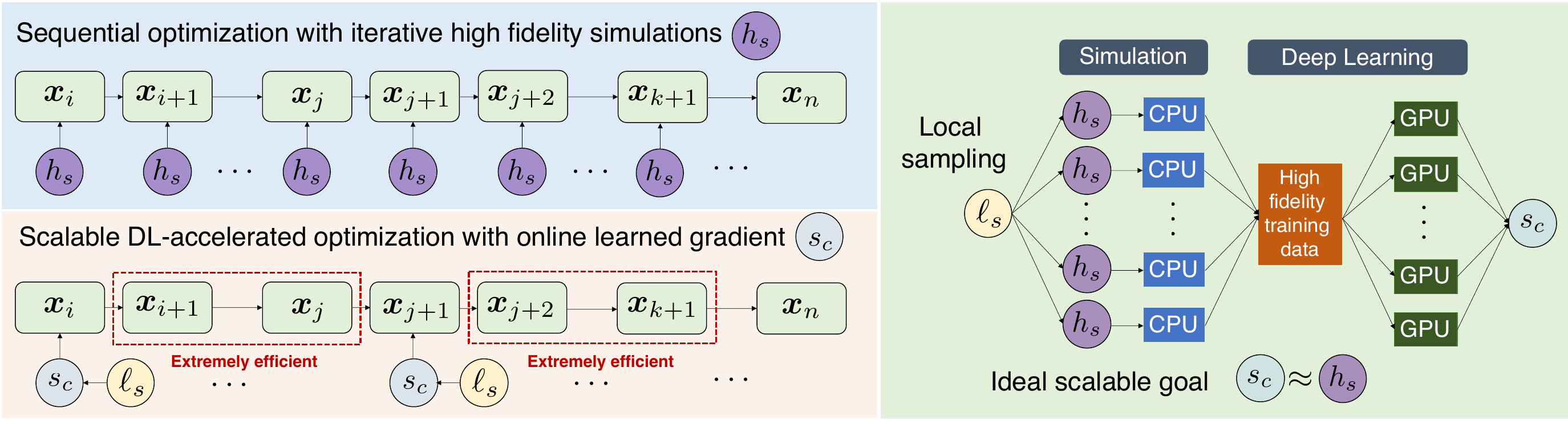}
    \caption{Scalable implementation of the SDL-TO framework on multiple CPUs and GPUs}
    \label{fig:parallel}
\end{figure}

{\bf Scalable implementation of the SDL-TO framework} \quad A core contribution of the SDL-TO framework is to utilize high performance computing (HPC) sources (multiple CPUs and GPUs) to accelerate the sequentially iterative TO process in a parallel scheme. Figure \ref{fig:parallel} shows the schema of our scalable implementation. Compared with the sequential TO iterative process $\bm x_i, \bm x_{i+1}, ..., \bm x_{n}$ with high fidelity FEM simulation $h_s$ at each step, we accelerate optimization with an online update using the learned gradient $s_c$ incorporating with a local sampling strategy $\ell_s$, where local samples of $h_s$ are generated from a Gaussian distribution in Eq.~\eqref{eq:gaussian} and sent to multiple CPUs for parallel simulations. Then we collect all $h_s$ data together for the distributed deep learning training on multiple GPUs to obtain a fast and accurate mapping between the current design and its gradient. In such a way, we only need to call a very small number of {\textit{learning}} steps, but use the learned gradient $s_c$ for a fast design update. For example, we call the learning step at $\bm x_{j+1}$ and then we can use the learned gradient for multiple later updating steps $\bm x_{j+2}...,\bm x_{k+1} (k>j+1)$. Ideally, the total computational cost of $s_c$ equals $h_s$ but it needs to be fine-tuned and optimized due to the communication cost and a requirement of large enough computing resources. In this work, we use {\texttt{mpi4py}} for implementing the parallel simulation on multiple GPUs and {\texttt{Horovod}} \cite{sergeev2018horovod} for the distributed training on multiple GPUs. All the parallel scheme is conducted and tested on the ORNL OLCF Summit supercomputer, where each Summit node consists of 2 IBM Power9 CPUs, 6 NVIDIA V100 GPUs, and NVLink for CPU-CPU and CPU-GPU communications.

\section{Examples}

\begin{figure}[h!]
     \centering
     \includegraphics[width=0.98\textwidth]{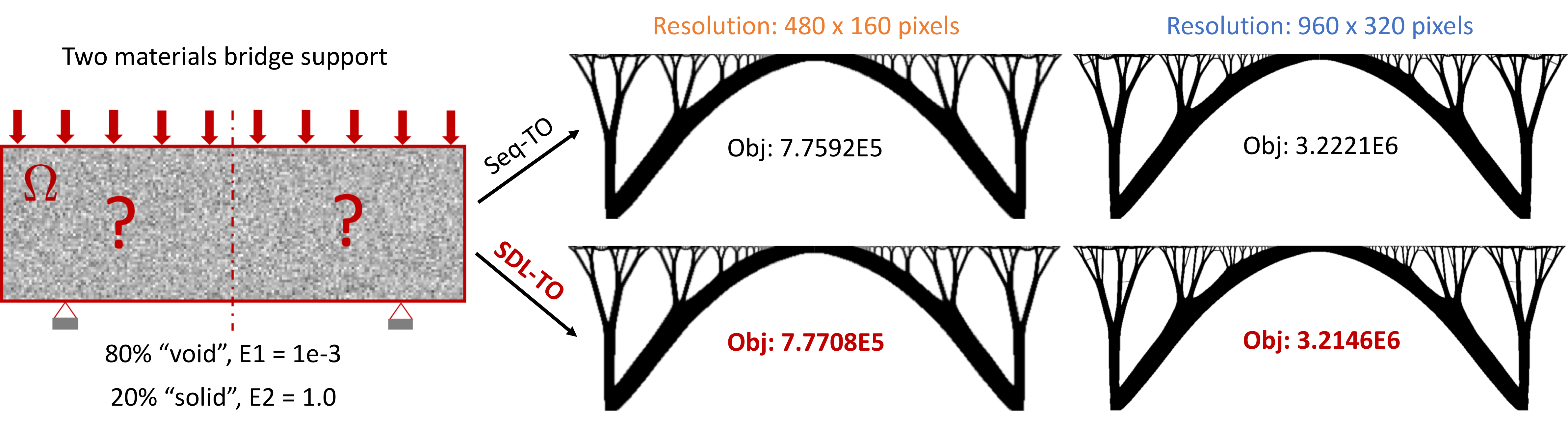} \quad 
    \caption{Comparing baselines on the bridge structure design example on two mesh grids (480$\times$160 pixels and 960$\times$320 pixels). We found that the SDL-TO shows competitive performance on both the objective loss and the final design compared to the sequential TO that uses the MMA optimizer.}
    \label{fig:bridge1}
\end{figure}

{\bf Benchmark bridge structure} \quad We first demonstrate our proposed framework on a benchmark bridge design problem. As shown in Figure \ref{fig:bridge1}, there are 80\% void materials with $E_1=0.001$ and 20\% solid materials with $E_2=1.0$ so that our goal is to optimize the layout of the materials to achieve minimum structural compliance of the support structure given a unit uniform load on the top. The SDL-TO framework shows the competitive performance of the objective loss values on two resolutions: 480$\times$160 pixels (mesh grids) and 960$\times$320 pixels (mesh grids). When the resolution is higher, our method outperforms the sequential TO (Seq-TO) method and shows smaller structures that benefit to improve the overall structural stiffness (the smaller objective (Obj) value, the better performance).

\begin{figure}[h!]
     \centering
     \includegraphics[width=0.98\textwidth]{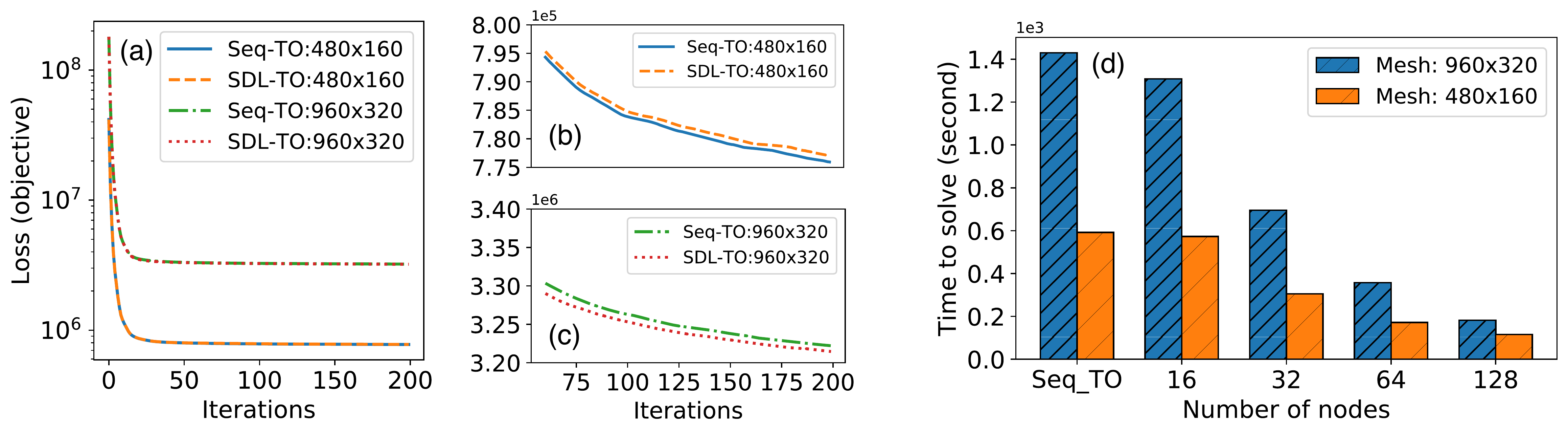} \quad 
    \caption{Quantitative comparison of the SDL-TO and the Seq-TO on objective loss and time-to-solution. (a) Iterative history of two mesh grids, zoom-in results of (b) 480$\times$160 pixels and (c) 960$\times$320 pixels, and (d) time-to-solution performance as the number of nodes increases.}
    \label{fig:bridge_all}
\end{figure}

The performance can be also demonstrated by a quantitative comparison of SDL-TO and Seq-TO in terms of iteration history and time-to-solution metric, as shown in Figure \ref{fig:bridge_all}. The objective (loss) values are very close in terms of the iteration process. To illustrate the small differences, Figure \ref{fig:bridge_all}(b) and (c) display the zoom-in results of two resolutions in the 75-200 iteration steps. Figure \ref{fig:bridge_all}(d) shows the scalability performance in terms of time-to-solution. The Seq-TO is compared with multiple numbers of nodes including 16, 32, 64, and 128. It is interesting to note that the SDL-TO framework shows stronger scalability in the high-resolution case than the low-resolution case. This is probably due to a higher ratio of the communication overhead in the low-resolution case.  

\begin{figure}[h!]
     \centering
     \includegraphics[width=0.95\textwidth]{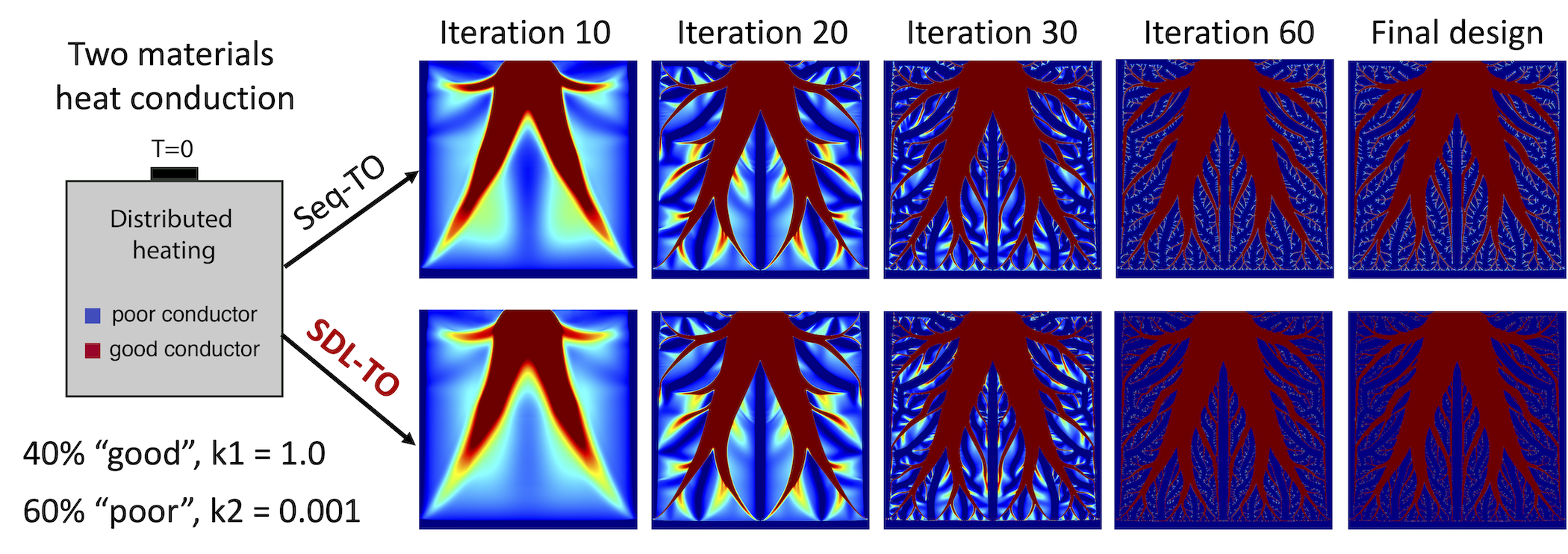} \quad 
    \caption{Two materials heat conduction design problem. Compared with the Seq-TO algorithm that uses 200 iterations, the SDL-TO demonstrates a very close performance but only using 12 gradient learning steps. The other updates are achieved by using the learned surrogate gradient.}
    \label{fig:heat1}
\end{figure}

{\bf Two materials heat conduction} \quad This is a large-scale TO problem that aims to address a challenging problem in additive manufacturing (AM): how to effectively distribute heating by optimizing the layout of two materials consisting of 40\% good conductor with $k_1=1.0$ and 60\% bad conductor with $k_2=0.001$, as shown in Figure \ref{fig:heat1}. Considering the AM constraints in practice, we use a very high resolution, 1024$\times$1024 mesh grids in this case for a large-scale demonstration. Thus, the conventional sequential TO process is more computationally intensive than the bridge design example discussed above. Here, we assume to use a total of 200 iterations for the optimization design and compare the performance of the final design and time-to-solution between the Seq-TO method and SDL-TO method. 

\begin{figure}[h!]
     \centering
     \includegraphics[width=0.8\textwidth]{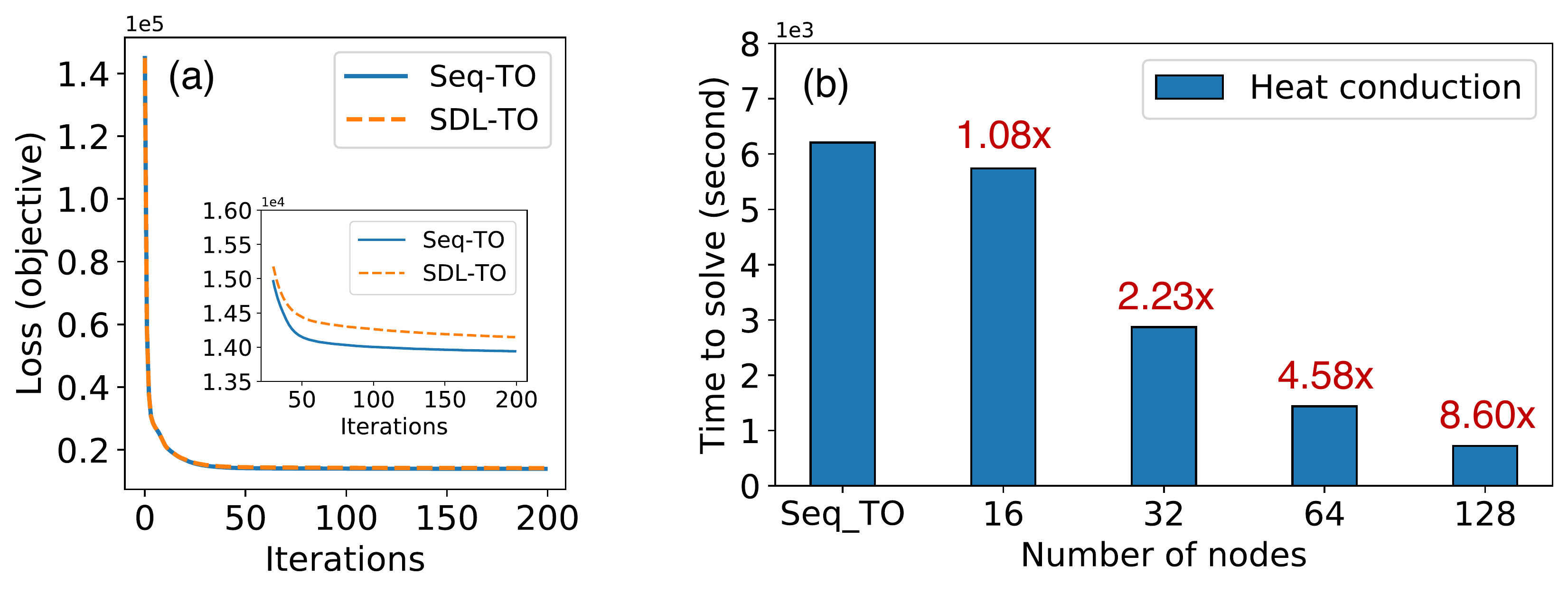} \quad 
    \caption{Iteration history and scalablility performance. The SDL-TO framework demonstrates a speed up of 8.6x accelerating over the standard TO implementation.}
    \label{fig_func}
\end{figure}

By comparison of the different iterative designs, specifically on iteration 10, 20, 30, and 60, as shown in Figure \ref{fig:heat1}, we can demonstrate that the SDL-TO framework using surrogate gradients learned by deep learning presents competitive performance on the final objective. For the iterative objective values shown in Figure \ref{fig_func}, there are also almost no differences between the sequential TO methods and the proposed scalable TO approach. However, the SDL-TO framework only uses 12 gradient learning steps and then uses the surrogate gradient for a fast design update. In terms of the time-to-solution performance, the SDL-TO method outperforms the sequential TO method. As shown in Figure \ref{fig_func}, the SDL-TO framework shows a speed up of 1.08x, 2.23x, 4.58x, and 8.60x using 16, 32, 64, and 128 nodes respectively. 

\section{Related work} 
{\bf Parameterization TO} \quad Hoyer et al. \cite{hoyer2019neural} proposed a neural reparameterization that improves topology optimization. They considered the use of implicit bias over functions by neural networks to improve the parameterization of TO and optimize the parameters of the neural networks instead of directly optimizing the densities on the mesh grid. This method can potentially achieve a better design with smaller objective loss values but it did not show how to accelerate the TO process within the proposed framework. Kallioras et al. \cite{kallioras2020accelerated} proposed a new representation using deep belief neural (DBN) coupled with the SIMP approach to accelerate the TO process. This method reduces the number of iterations but the overall time-to-solution is not ideal due to the relatively large expense of the DBN training and data generation.  

{\bf Neural networks and TO} \quad Several recent work focus on the use of convolutional neural network (CNN) \cite{banga20183d,zhang2019deep,sosnovik2019neural,yang2019topology,kollmann2020deep} or 
generative adversarial network (GAN) \cite{nie2020topologygan,yu2019deep,oh2019deep} to replace the standard TO methods. Most of them start from creating a dataset of structures/materials via the standard TO and then train a machine learning model based on the dataset. These methods can only reproduce their training data that needs a large computational cost in advance. The trained model is strongly limited by the training dataset and is difficult to handle a general problem. In contrast, our method pursues a general framework to accelerate the TO process by using the iterative history data rather than the end of the final design. 

\section{Conclusion}
In this work, we develop a scalable framework that combined deep learning and parallel computing to accelerate the TO process for designing large-scale additively manufactured materials. The novel contribution is to deeply understand the iterative history data and utilize deep neural networks (DNN) to learn an accurate and fast surrogate gradient instead of the true gradient that is often time-consuming. The learning process is achieved by utilizing parallel computing on multiple CPUs and a distributed DNN training on multiple GPUs, incorporated with a local sampling strategy that enables to capture of the low-dimensional search space. The proposed SDL-TO framework is demonstrated by benchmark examples and compared with the standard sequential TO baseline methods. In the large-scale AM materials design example, SDL-TO shows competitive performance but significantly reduces the computational cost by achieving a speed up of 8.6x over the standard TO implementation. 

\section{Acknowledgments}
This work was supported by the U.S. Department of Energy, Office of Science, Office of Advanced Scientific Computing Research (ASCR), Applied Mathematics program under contract ERKJ352, ERKJ369; and by the Artificial Intelligence Initiative at the Oak Ridge National Laboratory (ORNL). This work used resources of the Oak Ridge Leadership Computing Facility, which is supported by the Office of Science of the U.S. Department of Energy under Contract No. DE-AC05-00OR22725.
\bibliographystyle{plain}
\bibliography{reference, library,clay,CS,quasi}

\end{document}